\documentclass[aps,pra,floatfix,twocolumn,floatfix,showpacs]{revtex4}

\begin{document}

\title{Calculation of the hyperfine structure of the superheavy elements Z=119 and Z=120$^+$}

\author{T. H. Dinh, V. A. Dzuba, and V. V. Flambaum}

\affiliation{School of Physics, University of New South Wales,
        Sydney, 2052, Australia}
\date{\today}

\begin{abstract}

The hyperfine structure constants of the lowest $s$ and $p_{1/2}$
states of superheavy elements Z=119 and Z= 120$^+$ are calculated
using {\em ab initio} approach. Core polarization and
dominating correlation effects are included to all orders. Breit and
quantum electrodynamic effects are also considered. Similar
calculations for Cs, Fr, Ba$^+$ and Ra$^+$ are used to control the
accuracy. The dependence of the hyperfine structure constants on
nuclear radius is discussed.

\end{abstract}


\maketitle

\section{Introduction}

Study of the hyperfine structure of heavy and superheavy elements is
an important source of the information about nuclear structure of
these elements (see, e.g. Ref.~\cite{Sobelman,Armstrong}). Hyperfine structure
intervals are proportional to nuclear moments, such as magnetic
dipole moment, electric quadrupole moment, etc. The values of these
moments can be extracted from the comparison of the calculations and
the measurements. Apart from that, the hyperfine structure intervals
are sensitive to electric charge and magnetic moment distributions
within the nucleus. Parameters of these distributions can often be
extracted from the analysis of the hyperfine structure subject to
sufficient experimental data and the accuracy of the calculations.

The hyperfine structure analysis can be even more important for the
superheavy elements ($Z>100$) where sources of the information are
very limited. The study of the superheavy elements are motivated by
the hypothetical {\it island of stability} in the region 
$Z$=114 to $Z=126$ where shell closures are predicted (see, e.g., 
\cite{bender}). Elements up to $Z=118$, excluding $Z=117$, have been
synthesized (see, e.g.,  Refs.~\cite{hofmann,oganessian}) and evidence
for the naturally occurring element $Z=122$ has been reported
\cite{marinov}. 

The use of the hyperfine structure analysis is limited to nuclei with
odd number of protons or neutrons. The heaviest examples of such
nuclei which can be found in the literature as being already observed
include, e.g. $^{288}_{115}$Uup and
$^{289}_{114}$Uuq~\cite{Oganessian07}. There are numerous similar
examples for smaller $Z$. The information about nuclear magnetic
dipole and electric quadrupole moments, charge and magnetic moment
distribution for these elements is practically absent.  

Some standard approaches to the analysis of the hyperfine structure do
not work for very high $Z$. Consider, for example the
Fermi-Segr\'{e}~\cite{Armstrong} formula with the Casimir relativistic
factor. It expresses the hyperfine structure constant of the $s$-state
of an external electron via its wave function in the origin.
\begin{equation}
  A_s = const \times |\psi(0)|^2 F_{rel}(Z\alpha)(1-\delta)Ry,
\label{eq:As}
\end{equation}
where $\alpha$ is the fine structure constant, $Ry$ is Rydberg,
$\delta$ is the correction due to finite nuclear size and
$F_{rel}(Z\alpha)$ is the relativistic factor
\begin{equation}
  F_{rel}(Z\alpha) = \frac{3}{\gamma(4\gamma^2-1)}, \ \ \gamma =
  \sqrt{1-Z^2\alpha^2}.
\label{eq:Frel}
\end{equation}
Formulae (\ref{eq:As},\ref{eq:Frel}) are widely used in the hfs
analysis, however they fail at very high $Z$. It is easy to see that
the relativistic factor (\ref{eq:Frel}) turns to infinity at $Z
= 118.7$. Therefore, it is likely to overestimate the relativistic
corrections at smaller $Z$ as well. The reason for this is that the
formulae treat finite nuclear size as a small
correction. Hydrogen-like wave functions for a point-like nucleus are
used to calculate relativistic factor. However, it is known that the
finite nuclear size correction for superheavy elements is not small
and cannot be treated as a perturbation (see, e.g.~\cite{EPL}). 

A combination of an analytical and numerical approaches were recently used in
Ref.~\cite{Dihn} to study the dependence of the hyperfine structure intervals
on nuclear radius. A formula was suggested which is in good agreement with
accurate numerical calculations for $s$-states of atoms wit $Z<100$. However,
this formula also fails at higher $Z$. 

In this paper we don't use any analytical approaches but just perform accurate
numerical calculations of the hyperfine structure constants for superheavy
atoms. We demonstrate that the calculation which use finite-size
nuclei with realistic charge and magnetic moment distribution are very
similar to the calculations for lighter atoms. We consider elements
E119 and E120$^+$. The latter may have hyperfine structure if there is
an isotope with odd number of neutrons. Neither of these elements
have been synthesized yet. However, the ways of their production and
physics of their nuclei are discussed in literature~\cite{Smolanczuk}. These
elements are heavier than any known element. Therefore, if calculation
of the hyperfine structure brings no surprises for them one can expect
no surprises for lighter elements as well. Also these elements have
very simple electron structure with one external electron above closed
shells. Therefore, very accurate calculations are possible for the
elements. In our previous work~\cite{E119} we have calculated energy
levels of E119 and E120$^+$. Apart from some expected relativistic
effects like larger fine structure and stronger attraction of the
$s$-states to atomic core, the spectra of these superheavy elements is
very similar to the spectra of their lighter analogies, Fr, Ra$^+$, Cs
and Ba$^+$. We expect similar trend for the hyperfine structure and we
perform the calculations for the same set of atoms. This gives us an
estimate of the accuracy of the results for superheavy elements. We
stress that although the calculations bring no major surprise, the
dependence of the hyperfine structure on the nuclear radius is
significantly stronger for the superheavy elements than for their
lighter analogies. 

\section{Method of calculation}

We perform the calculations using a totally {\em ab initio} method developed
in our previous works \cite{E119,Dzuba84,Dzuba87,Dzuba89a,Dzuba89b,Dzuba2006}.
It starts from the relativistic Hartree-Fork (RHF) calculations for atomic
core and includes dominating correlation and all core polarization corrections
to all orders.  

Single-electron orbitals are found by solving a system of the RHF equations
for $N-1$ electrons of the closed-shell core (the $V^{N-1}$ approximation).
The RHF Hamiltonian has a form
\begin{equation}
\label{hf}
\hat H_{0}=c\mbox{\boldmath$\alpha$}\cdot{\bf p}+(\beta-1)m c^{2}+V_{\rm nuc}(r)+V^{N-1} \ .
\end{equation}
Here $\mbox{\boldmath$\alpha$}$ and $\beta$ are Dirac matrixes, 
$V_{\rm nuc}(r)$ is nuclear potential, 
$V^{N-1}=V_{\rm dir}+V_{\rm exch}$ is the sum of the direct and 
exchange Hartree-Fock potentials, $N$ is the number of electrons.
At distances much larger than nuclear radius $r_N$ nuclear potential is given
by $V_{\rm nuc}(r) = -Ze^2/r$, at short distances $V_{\rm nuc}(r)$ is obtained
by numerical integration of the Fermi distribution of nuclear electric
charge. We use $d=2.3$~fm as the thickness of the distribution and the
data from Ref.~\cite{Angeli} for the radii (see
Table~\ref{tab:iso}). We use $r_N = 1.1
(2.5Z)^{1/3}$~fm for superheavy elements.

The hyperfine interaction (HFI) is included in a self-consistent way as well.
The time-dependent Hartree Fock method (TDHF)~\cite{Dzuba84}, which is equivalent to
the well-known random-phase approximation (RPA) is used for this. 
To take into account finite nuclear size we use a simple model which
represents the nucleus as an uniformly magnetized ball. In our calculations
magnetic nuclear radius is the same as the electric one. However, these two
parameters can be varied independently.

The HFI Hamiltonian is given by
\begin{equation}
\label{hfe}
\hat H_{hfi}=e \mbox{\boldmath$\mu$}\cdot{\bf F(r)} ,
\end{equation}
\begin{equation}
\label{exter}
{\bf F(r)}=\left\{ \begin{array}{ll} \frac{\mbox{\bf r}\times 
\mbox{\boldmath$\alpha$}}{r_{m}^3}, & ~r < r_{m}~   \\*
\frac{\mbox{\bf r}\times\mbox{\boldmath$\alpha$}}{r^3},    
  & ~r \geq r_{m}~ \end{array} \right. \nonumber\ 
\end{equation}
where $r_m$ is the magnetic nuclear radius.

The TDHF equations have a form
\begin{equation}
\label{delpsi}
\hat (H_{0}-\epsilon_a)\delta\psi_a=(-F_{z}-\delta V^{N-1}+\delta\epsilon_a) \psi_a
\end{equation}
\begin{equation}
\label{delepsilon}
\delta\epsilon_a=\langle\psi_a|F_{z}+\delta V^{N-1}|\psi_a\rangle.
\end{equation}
Here index $a$ numerates states in the closed-shell core.
These equations are solved self-consistently for all states in the core.

States of the valence electron are calculated in the frozen field of atomic
core complemented by the {\em correlation potential} operator $\hat
\Sigma$~\cite{Dzuba87}
\begin{equation}
\label{Brueck}
  (\hat H_0 +\hat \Sigma -\epsilon) \psi_v^{BO} = 0.
\end{equation}
Here index $v$ numerates valence states. Correlation potential $\Sigma$
includes all lowest second-order correlation corrections and dominating
higher-order correlation corrections~\cite{Dzuba89a,Dzuba89b}. These higher-order
correlations include screening of Coulomb interaction and hole-particle
interaction. They are taken into account in all orders. Solving equations
(\ref{Brueck}) for valence states we find the so called {\em Brueckner}
orbitals for the valence states. This is emphasized by using superscript $BO$
for the orbitals. 

Total energy shift for the valence state $v$ due to HFI and correlations is
given by
\begin{equation}
\label{deltae}
\delta \epsilon_v = \langle \psi_v^{BO}| F_{z} + \delta V^{N-1} +
\delta \hat \Sigma | \psi_v^{BO} \rangle.
\end{equation}
Here $\delta \hat \Sigma$ is the change to the correlation potential $ \hat
\Sigma$ due to the hyperfine interaction. The term with $\delta \hat \Sigma$
is often called {\em structure radiation}. Finally, there is a
contribution due to the renormalization of the many-electron wave
function (see, e.g. \cite{Dzuba87})
\begin{equation}
\label{norm}
\delta\epsilon_{\rm norm} = -\langle \psi_v|F_z +\delta
V^{N-1}|\psi_v\rangle \langle \psi_v| \partial \hat \Sigma/\partial
E|\psi_v \rangle.
\end{equation}

Magnetic dipole hyperfine structure constant $A_v$ for the valence state $v$
is given by
\begin{equation}
\label{Av}
A_v = \frac{\mu e^2}{2m_pI}\frac{\delta \epsilon_v}{\sqrt{j_v(j_v+1)(2j_v+1)}}.
\end{equation}

\subsection{Breit and QED corrections}

It is hard to claim high accuracy of calculations for superheavy elements
without considering Breit and quantum electrodynamics (QED) corrections. 
We include Breit corrections in a very accurate way described in our previous
works \cite{Dzuba2006,Dzuba2001}. The QED correction are included
approximately via the QED potential suggested in Ref.~\cite{Flam05}.

The Breit operator has the form:
\begin{equation}
h^{B}=-\frac{\mbox{\boldmath$\alpha$}_{1}\cdot \mbox{\boldmath$\alpha$}_{2}+
(\mbox{\boldmath$\alpha$}_{1}\cdot {\bf n})
(\mbox{\boldmath$\alpha$}_{2}\cdot {\bf n})}{2r} \ ,
\end{equation}  
where ${\bf r}={\bf n}r$, $r$ is the distance between electrons, and 
$\mbox{\boldmath$\alpha$}$ is the Dirac matrix. It correspond to the zero
energy transfer approximation and includes magnetic interaction and
retardation. 

Similar to the hyperfine interaction, Breit operator induces a correction to
the self-consistent Hartree-Fock potential, which is taken into account in all
orders in Coulomb interaction by iterating the RHF equations with the
potential 
\begin{equation}
\label{VB}
V^{N-1}=V^{C}+V^{B} \ ,
\end{equation}  
where $V^{C}$ is the Coulomb potential, $V^B$ is the Breit potential.
The same potential (\ref{VB}) goes to the left and right-hand side of the TDHF
equations (\ref{delpsi}).

For the QED corrections we use a radiative potential derived in
Ref.~\cite{Flam05}. This potential was chosen to fit accurate calculations of
the QED corrections to the energies. It may give less
accurate results for the hyperfine structure. Therefore, we consider current
calculations of the QED corrections as rough estimations only.

\section{Results}

\begin{table}
\caption{Isotopes of Cs, Ba, Fr and Ra for which hyperfine structure
  constants have been calculated in present work. Magnetic moments are
in nuclear magnetons; $g_I \equiv \mu/I$.}
\begin{ruledtabular}
\label{tab:iso}
\begin{tabular}{cccccc}
  Isotope & $\mu$ & $I$ & $g_I$ & $r_N$[fm] & rms[fm]\footnotemark[1]\\
\hline
$^{133}$Cs & 2.582024 & 7/2 & 0.737721 & 5.671 & 4.8041 \\
$^{135}$Ba & 0.837943 & 3/2 & 0.558629 & 5.703 & 4.8273 \\
$^{211}$Fr & 4.00(8)  & 9/2 & 0.889    & 6.717 & 5.5545 \\ 
$^{225}$Ra & -0.7348(15) & 1/2 & -1.4696 & 6.887 & 5.6781 \\
\end{tabular}
\footnotetext[1]{Reference.~\cite{Angeli}}
\end{ruledtabular}
\end{table}

Table \ref{tab:iso} lists isotopes of lighter analogies of the
superheavy elements E119 and E120$^+$ for which the hyperfine
structure constants are calculated. The results of the calculations
are presented in Table~\ref{tab:Hyperfine}. Here RHF corresponds to
the $\langle \psi_v |F_z|\psi_v\rangle$ matrix elements with the
Hartree-Fock wave functions $\psi_v$, the RPA corresponds to the
$\langle \psi_v |F_z+\delta V^{N-1}|\psi_v\rangle$ matrix elements; BO
and RPA(BO) columns correspond to the same matrix elements but with
Hartree-Fock wave function replaced by Brueckner orbitals; the ``Str+Norm''
column includes structure radiation and renormalization.

\begin{table*}
\caption{Hyperfine structure constants of the lowest $s_{1/2}$ and
  $p_{1/2}$ states of Cs, Fr, E119, Ba$^+$, Ra$^+$, and E120$^+$ in
  different approximations in 
  MHz (Cs,Fr,Ba$^+$, Ra$^+$) and $g_I \times {\rm MHz}$ (E119, E120$^+$).}
\begin{ruledtabular}
\label{tab:Hyperfine}
\begin{tabular}{l l r r r r r r r r r}
 Atom & State & RHF & RPA & BO & RPA(BO) & Breit & Rad. &
 Str+Norm & Total & Exp.\\
\hline
Cs    & $6s$      & 1425 & 1718 & 1970 & 2325 & 6 & -21 & -31 & 2279 & 2298.2\footnotemark[1]\\ 
      & $6p_{1/2}$ &  161 &  202 &  240 &  294 & 0 &   0 &   5 &  299 & 291.89\footnotemark[1]\\

Fr    & $7s$      & 5791 & 6875 & 7716 & 8967 & 33 & -162 & -120 & 8718 & 8713.9\footnotemark[2]\\ 
      & $7p_{1/2}$ &  623 &  772 &  968 & 1180 & -4 &   -4 &    8 & 1180 & 1142\footnotemark[2]\\

E119  & $8s$      & 39344 & 46781 & 45531 & 53306 & 210 & -553 & -1315 & 51648 & \\ 
      & $8p_{1/2}$ &  5141 &  6165 &  8751 & 10506 & -65 &  -41 &  135 & 10535 & \\

Ba$^+$   & $6s$ & 2607 & 3095 & 3147 & 3684 & 8 & -42 & -82 & 3568 & 3591.6706(3)\footnotemark[3]\\ 
         & $6p_{1/2}$ & 441 & 530 & 568 & 674 & -1 &  0 & 4 &  677 & 664.2(2.4)\footnotemark[3]\\

Ra$^+$   & $7s$ & -21357 & -25022 & -25114 & -28986 & 92 & 436 & 668 & -27790 & -27684(13)\footnotemark[4] \\ 
         & $7p_{1/2}$ & -3626 & -4330 & -4746 & -5611 & 19 & 19 & -31 & -5604 & -5446(7)\footnotemark[4] \\

E120$^+$ & $8s$ & 74195 & 86640 & 80396 & 92884 & 352 & -837 & -2790 & 89609 & \\ 
         & $8p_{1/2}$ & 16883 & 19849 & 22286 & 26218 & -117 & -92 & 12 & 26021 & \\
\end{tabular}
\footnotetext[1]{Reference.~\cite{Cshfs}}
\footnotetext[2]{Reference.~\cite{Frhfs}}
\footnotetext[3]{Reference.~\cite{Bahfs}}
\footnotetext[4]{Reference.~\cite{Rahfs}}
\end{ruledtabular}
\end{table*}

As can be seen from the table the most important corrections are the
many-body corrections associated with the core polarization effect
(RPA) and with the correlation interaction of the external electron
with the core (BO). These effects follow approximately the same pattern
when moving from light to heavy atoms. This means that the accuracy of
the results should be about the same for all atoms and ions.

Breit contribution is small and can be neglected in all cases. This is
because Breit contributions are proportional to lower powers of $Z$
than other relativistic effects. The QED corrections are large for $s$
states. They reduce the hfs constants of these states by about 1\%. 

\begin{table}
\caption{Sensitivity of the hyperfine structure constants to the
  change of nuclear radius ($\kappa_{hr}$).}
\begin{ruledtabular}
\label{tab:r}
\begin{tabular}{cccccc}
Cs & Fr & \multicolumn{2}{c}{E119} & \multicolumn{2}{c}{E20$^+$} \\
$6s$ & $7s$ & $8s$ & $8p_{1/2}$ & $8s$ & $8p_{1/2}$ \\
\hline
-0.024\footnotemark[1] & -0.11\footnotemark[1] & -0.45 & -0.28 & -0.46 & -0.22 \\
\end{tabular}
\footnotetext[1]{Reference.~\cite{Dihn}}
\end{ruledtabular}
\end{table}

We also study the dependence of the hyperfine structure constants on
nuclear radius. This is done numerically by calculating the hfs
constants at different radius and then calculating the derivative
$dA/dr_n$ numerically. It is convenient to represent the results in a
form of the dimensionless constant $\kappa_{hr}$ as in
Ref.~\cite{Dihn}
\begin{equation}
  \kappa_{hr} = \frac{\delta A_v/A_v}{\delta r_n/r_n}.
\label{khr}
\end{equation}
Here $A_V$ is the hyperfine structure constant of the valence state
$v$, $r_n$ is nuclear radius. The results are presented in
Table~\ref{tab:r}. There are few things to note here. First, the
effect in superheavy elements is much larger than in their lighter
analogies. Second, the effect for $s$ and $p_{1/2}$ states is
significantly different. This represents an opportunity to use the
measurements of the hyperfine structure in superheavy elements not
only to extract nuclear magnetic moments but also to get some
information about nuclear radius. Note finally that the analytical
formulae describing the dependence of the hyperfine structure on
nuclear radius presented in Refs.~\cite{Sobelman,Dihn} do not work
here. They are not just inaccurate, they give absolutely meaningless
results. The reason for this is that the effect is large and cannot be
treated perturbatevely.

Calculating $\kappa_{hr}$ we assume that magnetic and electric radii
of the nucleus are the same. However, the program allows to treat
them independently and calculate two partial derivatives $\partial
A/\partial r_n$ and $\partial A/\partial r_m$, where $r_n$ is
electric radius and $r_m$ is magnetic radius. Such calculations show
that the hyperfine structure constants are more sensitive to the
change of the electric radius. Corresponding partial derivative is
approximately two time larger that those over the magnetic
radius. This is true for both $s$ and $p_{1/2}$ states.  

\section{Conclusion}

The hyperfine structure of lowest $s$ and $p_{1/2}$  states of the
superheavy elements Z=119 and Z=120$^+$ have been calculated  
with an uncertainty of a few percent. The dependence of the constants
on nuclear radius is presented.
The results may be used for experimental studies of nuclear,
spectroscopic and chemical properties of the elements.
 
\section*{Acknowledgment}

This research was supported in part by the Australian Research Council.


\begin{thebibliography}{99}
\bibitem{Sobelman}
        I. I. Sobelman, {\it Introduction to the Theory of Atomic Spectra},
        Nauka, Moscow, (1977) (Russian).
\bibitem{Armstrong}
        L. Armstrong, Jr., {\it Theory of the Hyperfine Structure of
          Free Atoms } (Wiley-Interscience, New York, 1971).

\bibitem{bender} M. Bender, P.-H. Heenen, and P.-G. Reinhard, Rev. Mod. Phys. {\bf 75}, 121 (2003).

\bibitem{hofmann} S. Hofmann and G. M\"unzenberg, Rev. Mod. Phys. {\bf 72}, 733 (2000).

\bibitem{oganessian} Y. Oganessian, Phys. Scr. T {\bf 125}, 57 (2006).

\bibitem{marinov} A. Marinov {\it et al.}, e-print arXiv:0804.3869 (2008).

\bibitem{Oganessian07} Y. Oganessian, Int. J. Mod. Phys. E {\bf 16}, 949 (2007).

\bibitem{EPL}  V. A. Dzuba and V. V. Flambaum,
       EPL {\bf 84}, 22001 (2008).

\bibitem{Dihn} T. H. Dinh, A. Dunning, V. A. Dzuba, and V. V. Flambaum,
       Phys. Rev. A {\bf 79} 054102 (2009).

\bibitem{Smolanczuk} R. Smolanczuk, Phys. Rev. C {\bf 63}, 044607 (2001).

\bibitem{E119} T. H. Dinh, V. A. Dzuba, V. V. Flambaum, and
  J. S. M. Ginges,  Phys. Rev. A {\bf 78}, 022507 (2008).

\bibitem{Dzuba84} V. A. Dzuba, V. V. Flambaum, and O. P. Sushkov, 
J. Phys. B  {\bf 17},  1953 (1984).

\bibitem{Dzuba87} V. A. Dzuba, V. V. Flambaum, P. G. Silvestrov, and O. P. Sushkov,  
  J. Phys. B {\bf 20}, 1399 (1987).

\bibitem{Dzuba89a} V. A. Dzuba, V. V. Flambaum,  and O. P. Sushkov,   
  Phys. Lett. A {\bf 141}, 147 (1989).

\bibitem{Dzuba89b} V. A. Dzuba, V. V. Flambaum, A. Ya. Kraftmakher, and O. P. Sushkov,   
  Phys. Lett. A {\bf 142}, 373 (1989).

\bibitem{Dzuba2006} V. A. Dzuba, V. V. Flambaum, and M. S. Safronova,
  Phys. Rev. A  {\bf 73}, 022112 (2006).

\bibitem{Angeli} I. Angeli, At. Data and Nuc. Data Tables, {\bf 87}
  185 (2004).

\bibitem{Dzuba2001} V. A. Dzuba, C. Harabati, W. R. Johnson, and M. S. Safronova,
  Phys. Rev. A  {\bf 63}, 044103 (2001).

\bibitem{Flam05} V. V. Flambaum, and J. S. M. Ginges, Phys. Rev. A {\bf 72}, 052115 (2005).


\bibitem{Cshfs} E. Arimondo, M. Inguscio, and P. Violino,
  Rev. Mod. Phys. {\bf 49} 31 (1977).

\bibitem{Frhfs} J. E. Sansonetti, J. Phys. Chem. Ref. Data, {\bf 36},
  497 (2007).

\bibitem{Bahfs} L.-J. Ma and G. zu Putlitz, Z. Phys. A {\bf 277}, 107
  (1976).

\bibitem{Rahfs} W. Neu {\em et al}, Z. Phys. D {\bf 11}, 105 (1989).


\end{thebibliography}
\end{document}